\documentclass[12pt,a4paper]{article}
\pdfoutput=1

\usepackage[utf8]{inputenc}
\usepackage[english]{babel}
\usepackage{amsmath}
\usepackage{amsfonts}
\usepackage{amssymb}
\usepackage{graphicx}
\usepackage[left=2cm,right=2cm,top=2cm,bottom=2cm]{geometry}
\usepackage{epstopdf}
\usepackage{float}
\usepackage{caption}
\usepackage{subcaption}

\def\a{\alpha}
\def\b{\beta}

\def\d{\delta}
\def\eps{\epsilon}           

\def\k{\kappa}                    

  \def\w{\omega}
\def\p{\pi}
\def\r{\rho}                                     

\def\co{{\cal O}}

\begin{document}
\begin{titlepage}
\begin{center}
\Large \bf Wilson Loops of Klebanov-Strassler like Wrapped Brane Models
\end{center}
\begin{center}
Michael Warschawski \footnote{pymw@swan.ac.uk}
\end{center}
\author{Michael Warschawski\\
  Department of Physics\\
  Swansea University\\
  Singleton Park\\
  Swansea\\
  SA2 8PP\\  
  United Kingdom\\
  \texttt{pymw@swan.ac.uk}}
\date{\today}
\vskip 0.2truein
 \vskip 4mm
 \begin{center}
\vskip 0.2truein
 \vskip 4mm
 \it{Department of Physics, Swansea University\\
 Singleton Park, Swansea SA2 8PP, United Kingdom.}
 \vskip 5mm

\vspace{0.2in}
 \end{center}
 \vspace{0.2in}
\begin{abstract}
We describe the behaviour of the Wilson loops for wrapped $D5$ systems. We start with the simplest such system possible and then add features to it bit by bit, and show how the Wilson loop is affected by them. This analysis led to the discovery of phase transitions. An interpretation why they occur is given and that knowledge is then used to construct systems with several phase transitions.
\end{abstract}
\end{titlepage}
\cleardoublepage
\tableofcontents
\cleardoublepage
\section{Introduction}
The $AdS/CFT$ correspondence \cite{hep-th/9711200} links a large N d-dimensional conformal field theory to a quantum  theory in $AdS_{d+1}$. More specifically, local operators of the conformal field theory are related to fields of the dual quantum gravity theory \cite{Gubser:1998bc}.

However, AdS/CFT and the more general gauge-gravity theories can be used to study non-local operators as well. One of the most interesting of such operators is the Wilson loop \cite{Wilson:1974sk}. It is accessible from the string theory side, which is computationally significant and physically, it provides a basis for gauge invariant gluonic operators. 

But even more important here, the potential of a quark-antiquark pair can be related to the VEV of a rectangular Wilson loop, whose sides are equal to the $q\bar{q}$ separation and $T\rightarrow \infty$. In the gauge theory the Wilson loop $W(\cal C)$ along a curve $\cal C$ is defined as
\begin{equation}
W({\cal C})\equiv \frac{1}{N_c} Tr\{ P[ e^{i\oint_{\cal C}A_\mu dx^\mu}]\}.
\end{equation}
where $N_c$ is the number of colors, $P$ represents the path ordering of the exponential and $A_\mu$ are the gauge fields. In this paper we will assume the gauge fields to be in the fundamental representation, however higher order representations are feasible \cite{Wilsonhigh}. On the string side of the correspondence we have \cite{Rey:1998ik,Maldacena:1998im}
\begin{equation} \label{wilson}
\left\langle \frac{}{}W({\cal C})\right\rangle= \int_{\partial F({\cal C})} {\cal D} F e^{- S[F]}
\end{equation}
where $F$ is used to denote all fields of the string theory and $\partial F$ their boundary values along $\cal C$.

We can approximate \eqref{wilson} using the steepest descent method. Here, the surface spanned by the strings ending on $\cal C$ and obeying the Nambu-Goto action $S_{NG}(F)$, is minimised. Now, that we have approximated the VEV of the Wilson loop and we know that we can relate the $qq$ potential to this VEV by $\left\langle \frac{}{}W({\cal C})\right\rangle \approx e^{-ET}$ we see that
 \begin{equation} \label{Energy}
  E_{qq}\approx \frac{S_{NG}}{T} 
  \end{equation}

Convinced of the importance of Wilson loops, we will compute its properties for backgrounds proposed in \cite{cascades}, deepening our understanding of those solutions. These sets of solutions, are very interesting as they relate two very fruitful extensions of the original $AdS/CFT$ correspondence. One of the extension is referred to as Klebanov-Strassler (KS) Models \cite{KS}. Here, a deformation is introduced that corresponds to an imbalance in the gauge groups of the field theory side of the correspondence. This puts the field theory in the mesonic branch \cite{Dymarsky:2005xt,Krishnan:2008}. 

The other extension is usually referred to as wrapped brane models \cite{hep-th/9803131,MN}. Here, we obtain the field theory through a higher dimensional field theory on which we perform a Kaluza-Klein compactification with twisting.

Our solution are KS like as they share the same IR and UV behaviour, and nontrivially relate to wrapped-brane models as they can be constructed from such a model by the addition of $N_f$ sources and applying U-Duality, a solution generating technique introduced in \cite{MM}.

The setup of the paper will be as follows. In Section 2, we discuss how to compute Wilson loops using Holography. In Section 3, we introduce the backgrounds of interest and study them using Wilson loops. Section 4 then deals with the occurence of phase transitions and explains why and how they occur. Section 5 tests our interpretation of the phase transitions and provides a technique for creating backgrounds with arbitrary many phase transitions. Section 6 then summarises the results obtained.

\section{General Theory} \label{Sonnenschein}
Here, I will discuss how to compute Wilson loops, as well as the potential energy of the $qq$-pair using the ideas from above for holographic theories. The following is based on \cite{Sonnenschein:1999if}.
Consider a 10d spacetime:
\begin{equation}
ds^2=-g_{\mu\mu} (\rho)(dx^\mu)^2+g_{\rho\rho}(\rho)(d\rho)^2+g_{\theta\psi}(\rho)dx^\theta dx^\psi
\end{equation}
as well as the Nambu-Goto Action
\begin{equation} \label{standardNG}
S_{NG} = \int d\sigma d\tau \sqrt{\det[\partial_\alpha x^\mu\partial_\beta x^\nu G_{\mu\nu}]}
\end{equation}
Now using the string gauge $\tau=t$ and $\sigma=x$ and not allowing the string to explore then internal space, we can transform \eqref{standardNG} to 
\begin{equation}
S_{NG} =T \int dx  \sqrt{f^2(\rho(x))+g^2(\rho(x))(\partial_x \rho)^2}
\end{equation}
Where 
\begin{equation}
f^2(\rho(x)) = g_{tt}g_{xx}      \qquad  g^2(\rho(x)) = g_{tt}g_{\rho\rho}\\
\end{equation}
Now, one can either solve the Euler-Lagrange equations of this system or transfer to the Hamiltonian picture and note that the Hamiltonian in the $x$ direction is a constant of motion as the Lagrangian is independent of $t$. Either way, one will derive the following equation of motion for the string:
\begin{equation}
\frac{d\rho}{dx}=\pm\frac{f(\rho)}{g(\rho)}\frac{\sqrt{f^2(\rho)-f^2(\rho_0)}}{f(\rho_0)}
\end{equation}
From here one can calculate that the length $L$ of the static string configuration connecting two quarks is
\begin{equation} \label{length}
L=\int dx =2\int_{\rho0}^\infty\frac{g(\rho)}{f(\rho)}\frac{f(\rho_0)}{\sqrt{f^2(\rho)-f^2(\rho_0)}}\,d\rho
\end{equation}
Now that we have all the ingredients we need, we can use equation~\eqref{Energy} to obtain $E(L)$, the potential energy of the quark-antiquark pair as a function of their seperation. This result is generally divergent, as we have to assume infinite quark masses in order to make the qq-pair static, which in turn allows us to obtain the rectangular Wilson loops needed. The infinite potential energy can be renormalized by subtracting the mass of the quarks $m_q =\int^\infty_0 g(\rho)d\rho$. The renormalized result is
\begin{equation} \label{energy}
E=f(\rho_0)L+2\int_{\rho_0}^\infty\frac{g(\rho)}{f(\rho)}(\sqrt{f^2(\rho)-f^2(\rho_0)}-f(\rho))\,d\rho -2\int_0^{\rho0}g(\rho)\,d\rho
\end{equation}
Now, the backgrounds \cite{cascades} we will focus on might  only be related to wrapped brane models in the technical sense discussed in the introduction, but due the fact that wrapped brane models are relatively simple to study, we will sketch the construction of our solutions of interest on their basis. We start from a generic wrapped brane model, modify it by applying the solution generating technique referred to as U-Duality (or rotation) and add $N_f$ sources. First, these sources are introduced naively, but then will be distributed according to profiles that become increasingly complex in order to correct more and more unphysical behaviours that will be discussed in more detail. At every stage we will evaluate equations~\eqref{length}~and~\eqref{energy} to illustrate the behaviour of Wilson loops discussed. Afterwards we will construct cases, where the change of E(L) to linear behaviour is started off by a phase transition. This is then generalised to obtain solutions that produce several phase transitions.

\section{Towards the background}
The wrapped brane models consist of  a metric $ds^2$, a dilaton $\Phi$ and an RR-three form $F_3$. To write down the background, we use $SU(2)$ left-invariant one-forms
\begin{equation}
\tilde{\w}_1= \cos\psi\, d\tilde\theta+\sin\psi\sin\tilde\theta
\,d\tilde\varphi\,,\quad
\tilde{\w}_2=-\sin\psi\, d\tilde\theta+\cos\psi\sin\tilde\theta\,
d\tilde\varphi\,,\quad \tilde{\w}_3=d\psi+\cos\tilde\theta\, d\tilde\varphi\,,
\end{equation}\label{su2}
and the vielbeins
\begin{align}
&E^{x^i}= e^{\frac{\Phi}{4}}dx^i\,    ,\qquad E^{\r}=  e^{\frac{\Phi}{4}+k}d\r\,  ,\qquad E^{\theta}=  e^{\frac{\Phi}{4}+h}d\theta\,  ,\qquad E^{\varphi}= e^{\frac{\Phi}{4}+h} \sin\theta\, d\varphi \,, \\
&E^{1}=  \frac{1}{2}e^{\frac{\Phi}{4}+g}(\tilde{\omega}_1 +a\, d\theta)\,  ,\;\; E^{2}=\frac{1}{2}e^{\frac{\Phi}{4}+g}(\tilde{\omega}_2 
-a\,\sin\theta\, d\varphi)\,,\;\; E^{3}= \frac{1}{2}e^{\frac{\Phi}{4}+k}
(\tilde{\omega}_3 +\cos\theta\, d\varphi)\,. 
\end{align}
Thus, in the Einstein frame, we have
\begin{equation}
\begin{aligned}
ds_E^2	&= \sum_{i=1}^{10} (E^{i})^2\,,    \label{f3old}\\
F_3			&= e^{-\frac{3}{4}\Phi}\Big(f_1 E^{123}+ f_2 E^{\theta\varphi 
3}
+ f_3(E^{\theta23}+ E^{\varphi 13})+ 
f_4(E^{\r 1\theta}+ E^{\r\varphi 2})   \Big)\,, 
\end{aligned}
\end{equation}
where we defined
\begin{equation}
\begin{split}
&E^{ijk..l}=E^{i}\wedge E^{j}\wedge E^{k}\wedge...\wedge E^{l}\, ,\\
&f_1=-2 N_c e^{-k-2g}\,,\qquad\qquad f_2= \frac{N_c}{2}e^{-k-2h}(a^2 -
2 a b +1 )\,,\\
&f_3= N_ce^{-k-h-g}(a-b)\,,\qquad f_4=\frac{N_c}{2}e^{-k-h-g}b' \,.
\end{split}
\end{equation}
\subsection{Sourceless Wrapped-D5}
To connect to Section~\ref{Sonnenschein}, we have to move from the Einstein frame to the string frame. The metric can be obtained by multiplying $ds_E^2$ by $e^{\frac{\Phi}{2}}$. The metric elements relevant for the Wilson loop computation become
\begin{equation}
g_{tt}=e^{\Phi(\rho)}\quad g_{xx}=e^{\Phi(\rho)}\quad g_{\rho\rho}= e^{\Phi(\rho) + 2k(\rho)}
\end{equation}
There are 6 functions in the background $(a,b,\Phi,g,h,k)$ which depend on the radial coordinate $\rho$. The background is determined by solving the equations of motion for these functions. A system of BPS equations can be derived\cite{Casero:2006pt,HoyosBadajoz:2008fw,Casero:2007jj}. These equations show that all background functions can be expressed in terms of $Q(\rho)$ and $P(\rho)$, where
\begin{equation}
Q(\r)= N_c\,(2\r \coth(2\r)-1)
\end{equation}
and $P$ is a solution to the following differential equation
\begin{equation} \label{unfmaster}
 P'' + P'\Big(\frac{P'+Q'}{P-Q} +\frac{P'-Q'}{P+Q} - 4 
\coth(2\rho)
\Big)=0
 \end{equation} 
\eqref{unfmaster} is usually referred to as the master equation. The background functions of interest to us are
\begin{equation}\label{unfrel}
e^{2k}=\frac{P'}{2}\quad e^{4\Phi-4\Phi_0}=\frac{2\sinh(2\r)^2}{(P^2-Q^2)P'}
\end{equation}
A completely analytical solution is given by
\begin{equation}\label{vanilla}
P=2N_c\r.
\end{equation}
It leads to the background discussed in \cite{MN}.
Other solutions can be found semi-analytically. Interesting and well studied-solutions can be found by numerically connecting the large $\r$ expansion
\begin{eqnarray} \label{UV}
&P&=e^{4\rho/3}\Big[ c_+  
+\frac{e^{-8\r/3} N_c^2}{c_+}\left(
4\r^2 - 4\r +\frac{13}{4} \right)+ e^{-4\r}\left(
c_- -\frac{8c_+}{3}\r \right)+\nonumber\\
& & + \frac{N_c^4 e^{-16\r/3}}{c_+^3}
\left(\frac{18567}{512}+\frac{2781}{32}\r +\frac{27}{4}\r^2 +36\r^3\right)  + \co(e^{-20\rho/3})
\Big]\,
\end{eqnarray}
with the IR expansion
\begin{equation} \label{IRunf}
P= h_1 \r+ \frac{4 h_1}{15}\left(1-\frac{4 N_c^2}{h_1^2}\right)\r^3
+\frac{16 h_1}{525}\left(1-\frac{4N_c^2}{3h_1^2}-
\frac{32N_c^4}{3h_1^4}\right)\r^5+\co(\r^7)\,.
\end{equation}
See for example \cite{Elander:2011mh}. Here it was shown that this type of solution corresponds to the addition of a an irrelevant operator which in turn requires a UV completion.

Such a completion that again make contact with Klebanov-Strassler theories in the IR is given by a U-duality as it is described in \cite{MM}.This procedure has also been referred to as a rotation. It is a solution generating technique that schematically unites the backgrounds, by taking a solution to our master equation and mapping it to another background where new fluxes are turned on. While neither the background functions, nor $Q$ and $P$ change, the metric does and new fluxes are turned on. the metric elements relevant to our calculations are
\begin{equation}
g_{tt}=\frac{e^{\Phi(\rho)}}{\sqrt{\hat{h}}}\quad g_{xx}=\frac{e^{\Phi(\rho)}}{\sqrt{\hat{h}}}\quad g_{\rho\rho}= e^{\Phi(\rho) + 2k(\rho)}\sqrt{\hat{h}}
\end{equation}
where $\hat{h}=1-\k^2e^{2\Phi}$ and $ \k $ can be chosen to be $\k = e^{-\Phi(\infty)}$, see \cite{GMNP}.

We study the rectangular Wilson loop for \eqref{vanilla}, as well as for rotated and unrotated numerical solutions connecting \eqref{IRunf} with \eqref{UV}. In the numerical case, we first solve the master equation using \eqref{IRunf} and its derivative as initial conditions. We then obtain the background functions using the relations~\eqref{unfrel}. We then solve the Nambu-Goto Action, as outlined in Section~\ref{Sonnenschein} to obtain $L(\r_0)$ and $E(\r_0)$ given by \eqref{length} and \eqref{energy}. These two functions can be combined to give us $E(L)$. This leads to the following plots:
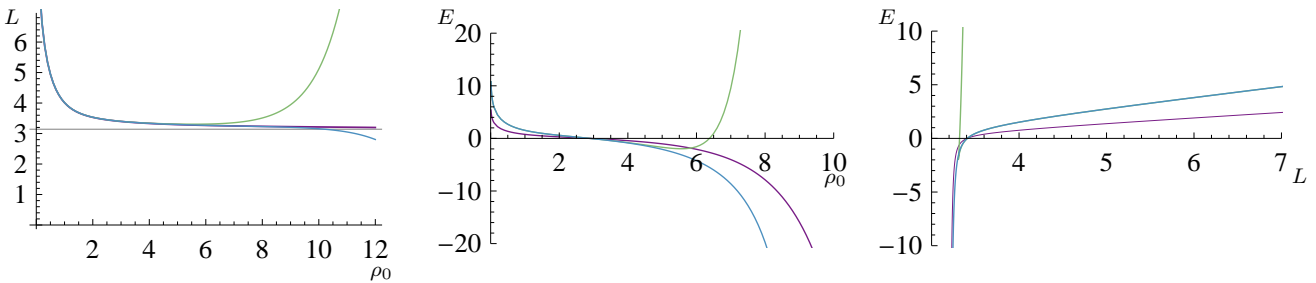
\begin{figure}[H]
\begin{center}
\begin{picture}(220,170)
\put(-140,0){\includegraphics[width=\textwidth]{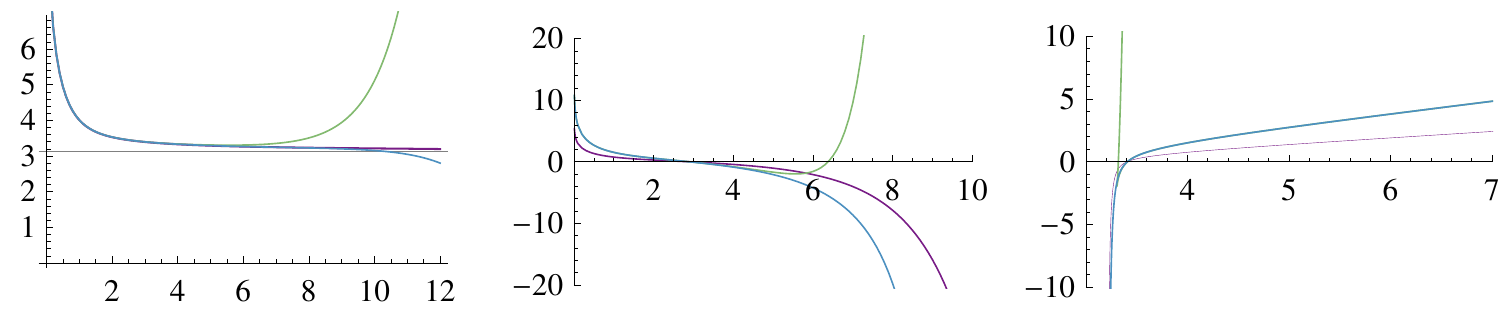}}
\put(-137,95){{\scriptsize{$L$}}}
\put(0,0){{\scriptsize{$\r_{0}$}}}
\put(25,95){{\scriptsize{$E$}}}
\put(170,35){{\scriptsize{$\r_{0}$}}}
\put(190,95){{\scriptsize{$E$}}}
\put(345,35){{\scriptsize{$L$}}}
\end{picture}
\caption{Purple represents $P=2N_c\r$, while green and blue are unrotated/rotated numerical solutions respectively. For the constants we chose $N_c=1$, $\Phi_0=0$ and $h_1=2+10^{-10}$}
\label{Fig:MN}
\end{center}
\end{figure}

For the solution with $P=2 N_c \r$, given by \eqref{vanilla}, we can see that there is a minimum seperation length beyond which the quarks cannot be forced closer to each other. In fact, we can easily calculate that 
\begin{align}
L(\r_0 \rightarrow \infty) = \pi \lim_{\r_0\rightarrow\infty} \frac{g(\r_0)}{\partial_{\r_0} f(\r_0)} = \pi.
\label{LinfMN}
\end{align}
This seems to indicate that the quarks are not point-like objects, but have a finite size. This is not very surprising as it is well known that the analytical solution is not dual to a Field Theory, but rather a Little String Theory.

The dynamics of the numerical solutions mirror $P=2N_c\r$ in the IR. In all three cases $E(L)$ behaves Coulomb like below the confinement scale, while $L(\r_0)$ initially asymptotes $\p$. This is expected, as $P$ is always linear below the confinement scale, but for the numerical solutions it then changes its behaviour to become exponential, in the order to match the UV expansions \eqref{UV}. At this point $L(\r_0)$ starts to increase exponentially for the unrotated case. This is due to the aforementioned irrelevant dimension 8 operator, coupling to gravity and thus causing the unphysical behaviour observed in the UV, where $E(L)$ grows exponentially. After a UV completion given by rotation, we see that  $L(\r_0)\rightarrow 0$ when $\r_0 \rightarrow \infty$ and we recover linear behaviour for $E(L)$ after the confinement scale, as expected for confining theories.

The analysis in this subsection has been very similar to \cite{Elander:2011mh}. However, in that paper, solutions to the master equation~\eqref{unfmaster} have been studied, for which it was assumed that $P\gg Q$. This leads to an IR behaviour of P that is constant to first order and not linear \cite{Nunez:2008}. This has led to some interesting effects like phase transitions. For our IR expansion~\eqref{IRunf} we have not been able to find any phase transitions. However, our case does never feature cusps in the string configuration that would break down the steepest descent approximation, either. 

\subsection{Flavor branes and Flavor profiles} \label{flavor}
First we add sources in the original, naive way as described in \cite{Casero:2006pt,arXiv:1002.1088}, without any kind of profile describing the distribution of the $N_f$ branes. This time the form of the metric remains unchanged, while the background functions follow different equations. See \cite{GMNP,Martucci:2005ht,Bigazzi:2005md,Casero:2006pt} for reference. $Q$ and $P$ are now given by
\begin{eqnarray}
& & P''+(P'+N_f )\Big[\frac{P'+Q'+2 N_f }{P-Q}
+ \frac{P'-Q'+2 N_f }{P+Q}
- 4 \coth(2\r)   \Big]=0\,,\nonumber\\
& & Q(\r)= \frac{2N_c-N_f}{2}(2\r \coth(2\r)-1)\,
\end{eqnarray}
We now have for our background functions
\begin{equation}
e^{2k}=\frac{P'+N_f}{2}\quad e^{4\Phi-4\Phi_0}=\frac{2\sinh(2\r)^2}{(P^2-Q^2)(P'+N_f)}
\end{equation}
We require new IR asymptotes reflecting the addition of $N_f$ sources. The ones befitting our case are \cite{GMNP}
\begin{equation}
P(\r)=h_1\r+\frac{4N_f}{3}\left(-\r\log\r-\frac{1}{12}\r\log(-\log\r)+\co\left(\frac{\r\log (-\log \r)}{\log \r}\right)\right)
+\co(\r^3\log \r)
\end{equation}
Unfortunately, we  have adopted an IR singularity caused by the high density stacking of the branes near $\r=0$ \cite{arXiv:1002.1088}. This leads to an unphysical background, which can also be seen through the study of the Wilson loops \cite{Bennett:2011xd}
\begin{figure}[H] 
  \centering
    \includegraphics[width=\textwidth]{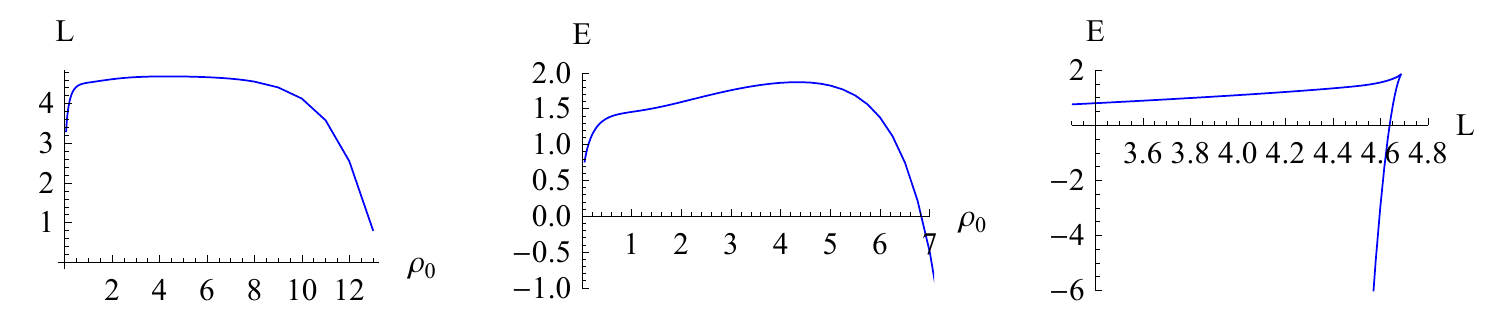}
    \caption{$\Phi_0=0$, $N_c=1$, $N_f=1$ and $h_1=4$}
\end{figure}
Note the cusp in the function $E(L)$. This is a visual indication of the fact that the way we introduced sources causes us problems. 

As described in \cite{Conde:2011rg}, we can avoid this singularity by distributing the sources such that not all sources reach $\r = 0$. Of importance is the fact that the density profile for the sources has a Maclaurin series with leading term at least of order $\co(\rho^3)$. A simple profile that fulfils the requirements - see \cite{Barranco:2011vt} - is
\begin{equation} \label{prof1}
S(\r)=\tanh^4(2\r).
\end{equation}
Note that its IR expansion is
\begin{equation}
S(\r)=16\r^4-\frac{256}{3}\r^6+\frac{1536}{5}\r^8+\co[\r^{10}]
\end{equation}

as required. Following \cite{Conde:2011rg}, we now have
\begin{align}
 & (P''+N_f S')+(P'+N_f S )\Big[\frac{P'+Q'+2 N_fS }{P-Q}
+ \frac{P'-Q'+2 N_fS }{P+Q}
- 4 \coth(2\r)   \Big]=0\\
& Q = \coth(2\r)\Big[\int_0^\r dx 
\frac{2N_c -N_fS(x)}{\coth^2(2x)}\Big]\nonumber\\
& e^{2k}=\frac{P'+N_fS}{2}\nonumber\\
&e^{4\Phi-4\Phi_0}=\frac{2\sinh(2\r)^2}{(P^2-Q^2)(P'+N_fS)}\nonumber
\end{align}
Please note, comparing with the case without flavor profile, it seems that the analysis was simply generalised by replacing $N_f \rightarrow N_f S(\r)$ everywhere. This is not quite true, seen for example at the new form the equation governing Q takes.

We get
\begin{figure}[H] 
  \centering
    \includegraphics[width=0.5\textwidth]{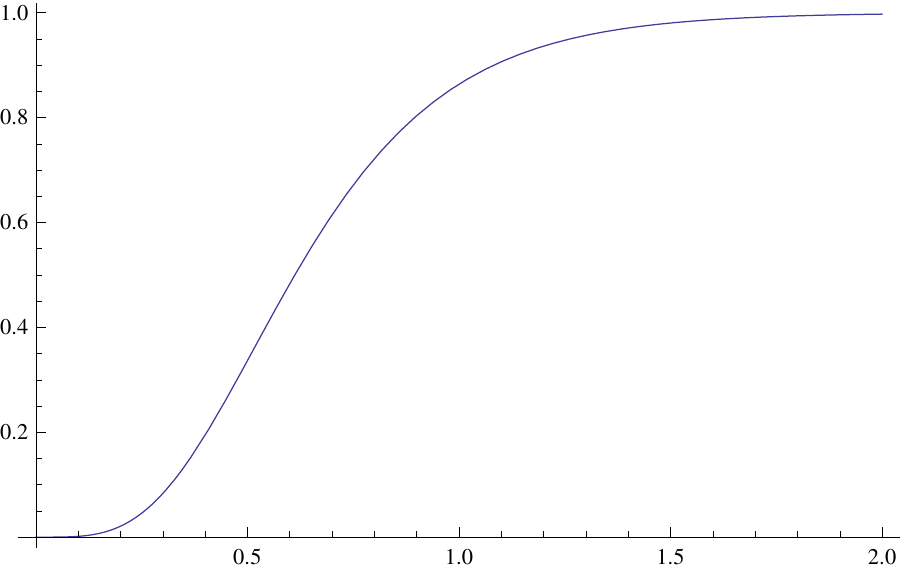}
    \caption{$S(\r)$}
\end{figure}

\begin{figure}[H] 
  \centering
    \includegraphics[width=\textwidth]{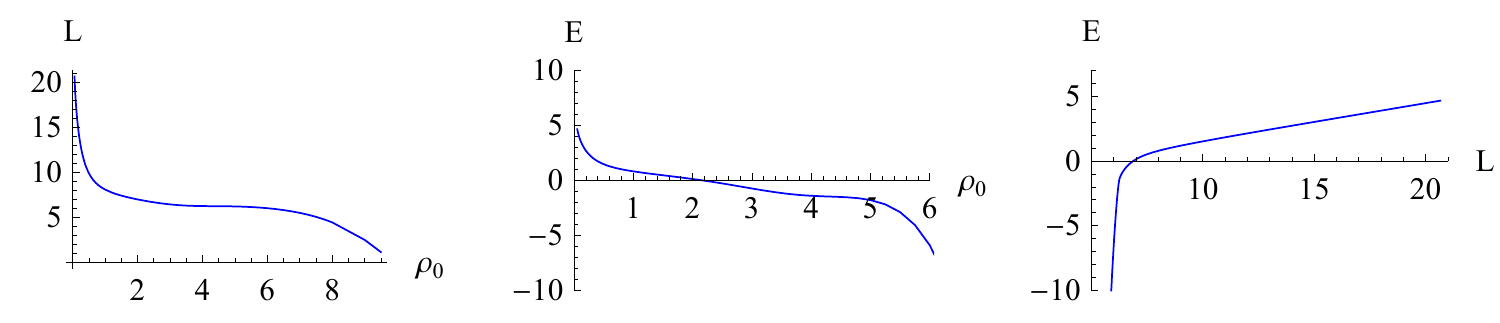}
    \caption{$\Phi_0=0$, $N_c=5$, $N_f=2$ and $h_1=11$}
\end{figure}

As one can see, $E(L)$ behaves smoothly again. Let us define $h_{1c}$ to be the value of $h_1$ for which $P$ remains linear in the UV. Note that $h_1 \approx h_{1c}$ here, thus $P$ remains linear for quite some time, resulting in $L(\r_0)$ plateauing before approaching $0$.
\subsection{Bump-like profiles} \label{bump}
As argued in \cite{cascades}, the UV behaviour is improved by using a profile that decays like $e^\frac{-4\r}{3}$. This is due to the fact that to preserve the KS like UV asymptotic behaviour, we need a profile that decays at least that fast. The presence of an exponentially increasing number of source branes, as is the case for a profile where $S \rightarrow 1$, behaves like the insertion of an irrelevant operator that deforms the UV dynamics. We also would not want a profile that decays faster to keep $T^{\textrm{sources}}_{x_0x_0}$, representing the mass density of the sources, positive everywhere. To have $T^{\textrm{sources}}_{x_0x_0} \rightarrow 0$ exactly, a profile that decays like $(\sinh(4\r)-4\r)^{-1/3}$ is needed\footnote{I thank Jerome Gaillard and Carlos N\'u\~nez for sharing that information prior to publication.}. Thus the following two profiles will be of importance. See \cite{Conde:2011rg,Barranco:2011vt} for more details.
\begin{equation} \label{prof2}
S(\r)=\tanh^4(2\r) e^\frac{-4\r}{3}\\
\end{equation}
\begin{equation} \label{prof3}
\hat{S}(\r)=\frac{\tanh^4(2\r)}{(\sinh(4\r)-4\r)^{1/3}}
\end{equation}
which look like
\begin{figure}[H] 
  \centering
    \includegraphics[width=0.5\textwidth]{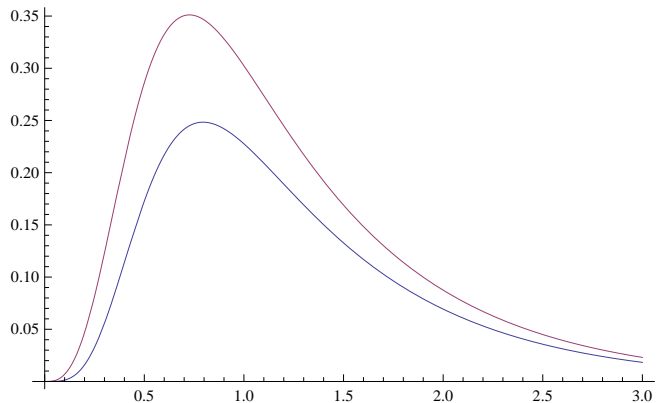}
    \caption{$S(\r)$ is blue while $\hat{S}(\r)$ is red}
\end{figure}
Note that $\hat{S}(0)=\frac{0}{0}$. While this can be quite easily analytically continued to be $\hat{S}(0)=0$, numerical calculations using this profile have to be dealt with carefully. The approach in this paper is based on the following idea. As the IR expansion of this profile is
\begin{equation}
\hat{S}(\r)=6^\frac{1}{3}(4\r^3-\frac{112}{5}\r^5+\co[\r^7])
\end{equation}
we define $\hat{S}(\r)=a\r^3+b\r^5$ for $0< \r < \epsilon$. We then solve for $a$ and $b$ by demanding that $\hat{S}(\r)$ and $\hat{S}'(\r)$ are continuous at $\epsilon$.
For simpler numerics at this stage, let us use $S(\r)$ for now, and observe that we get a very similar picture to the previous subsections.
\begin{figure}[H] 
  \centering
    \includegraphics[width=\textwidth]{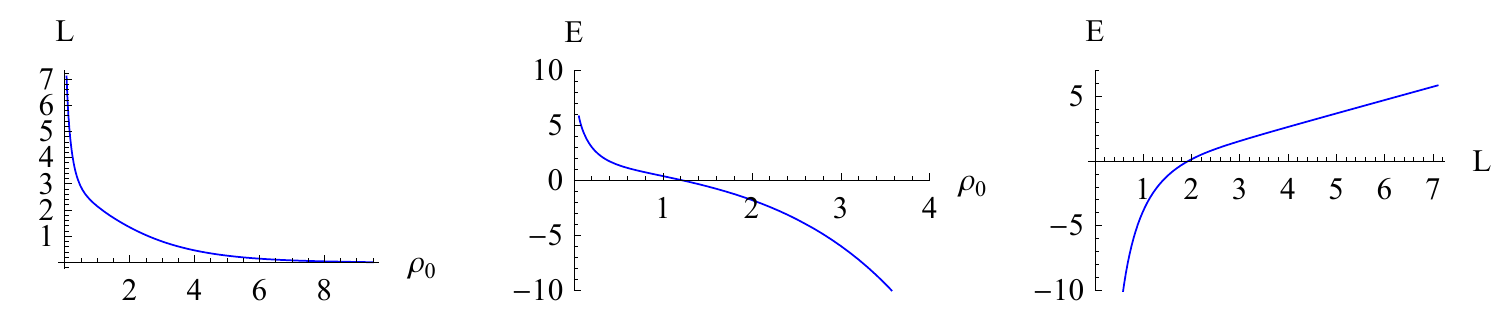}
    \caption{$\Phi_0=0$, $N_c=1$, $N_f=1$ and $h_1=3.12932$}
\end{figure}
As we can see, the behaviour of the Wilson loop is as expected for a confining theory, and no area of concern can be found due to the study of this observable in this background.
Note that, as described in \cite{cascades}, all profiles used in this paper lack a rigorous proof of existence, as the profiles are not derived from a kappa-symmetric embedding. However, the healthy behaviour of the backgrounds found in this and other works on these backgrounds suggests that they are nonetheless physically relevant.

\section{Phase Transition}
An interesting observable phenomenon that can be found is a phase transition. This was first observed for SQCD-like field theories in \cite{Bigazzi:2008gd}. The interpretation of such a phenomena is similar to the phase transition of a boiling Van-der-Waals gas. Several other papers have found such a phenomena in similar cases since. See for example \cite{walking,Conde:2011rg,Bennett:2011xd,Niall,Bigazzi:2008}

In order to find them for the theory discussed here, the parameter space of varying $h_1$ was explored. Please note that not all values of $h_1$ are valid. There always exists a critical value for $h_1$, For which $P$ grows linearly. For $h_1>h_{1c}$, $P$ grows exponentially in the UV and for $h_1<h_{1c}$, $P$ dies down to a singularity. Note the following figure. 
\begin{figure}[H] 
  \centering
    \includegraphics[width=0.5\textwidth]{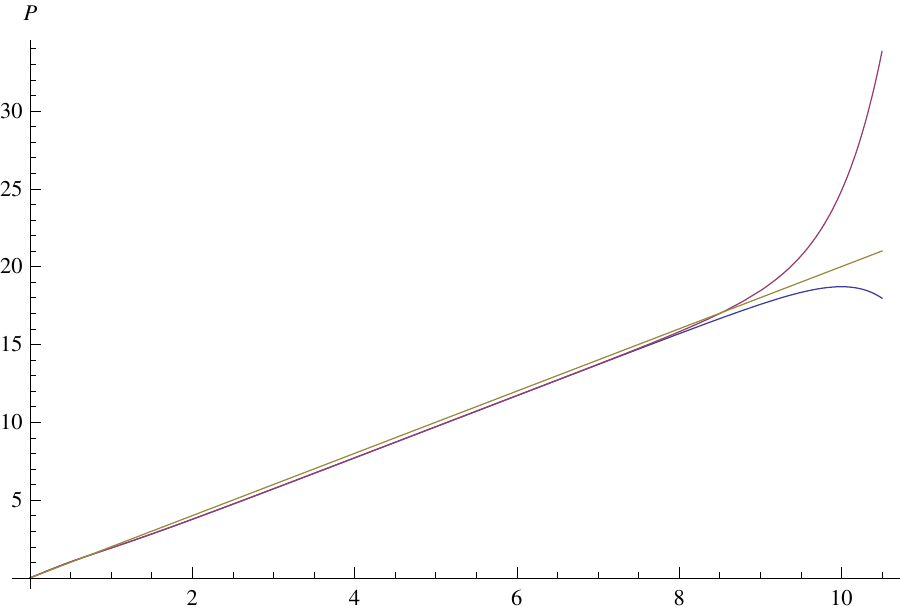}
    \caption{The green line represents $P$ for $h_{1c}$, red $h_{1c}+ \eps $ and blue $h_{1c}- \eps $}
    \label{P}
\end{figure}
Our requirement for $\Phi$ to be bounded only allows exponentially growing solutions like the red curve. Thus a shooting procedure was used to find $h_{1c}$, which was used as starting value for our exploration into ever increasing values of $h_1$. Unfortunately the search for a phase transition was unsuccessful for the three profiles \eqref{prof1}, \eqref{prof2} and \eqref{prof3} mentioned above.

To see how to proceed from here, let us analyse that Van-der-Waals gas analogy from the beginning of the section more closely. Such a gas follows an equation of state of the form 
\begin{equation}
P= \frac{NRT}{V- b N} - \frac{N^2 a}{V^2}
\end{equation}
where $R, b, a$ are constants. Look at the diagrams borrowed from \cite{walking}
\begin{figure}[H]
\centerline{\includegraphics[width=7cm]{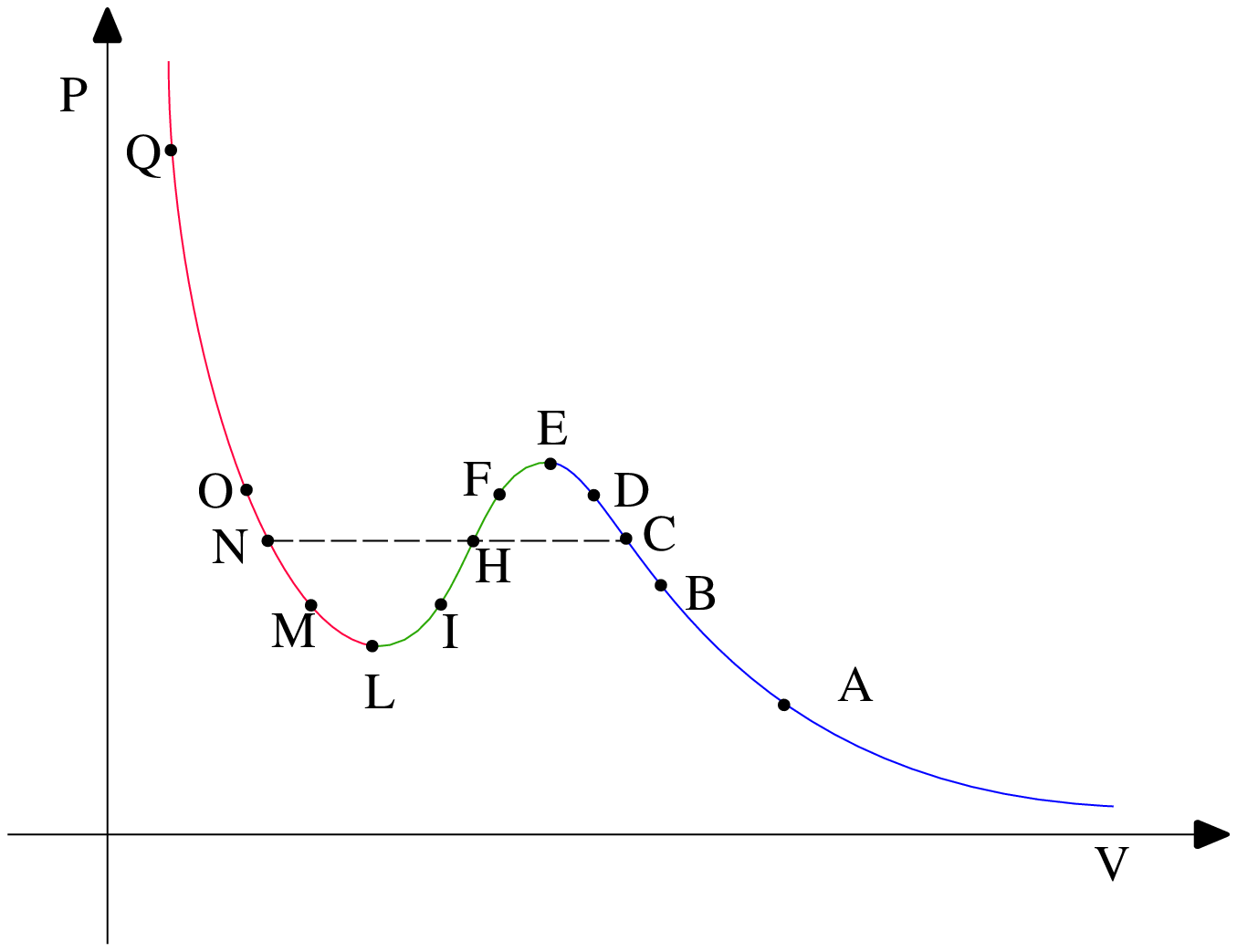}
\includegraphics[width=7cm]{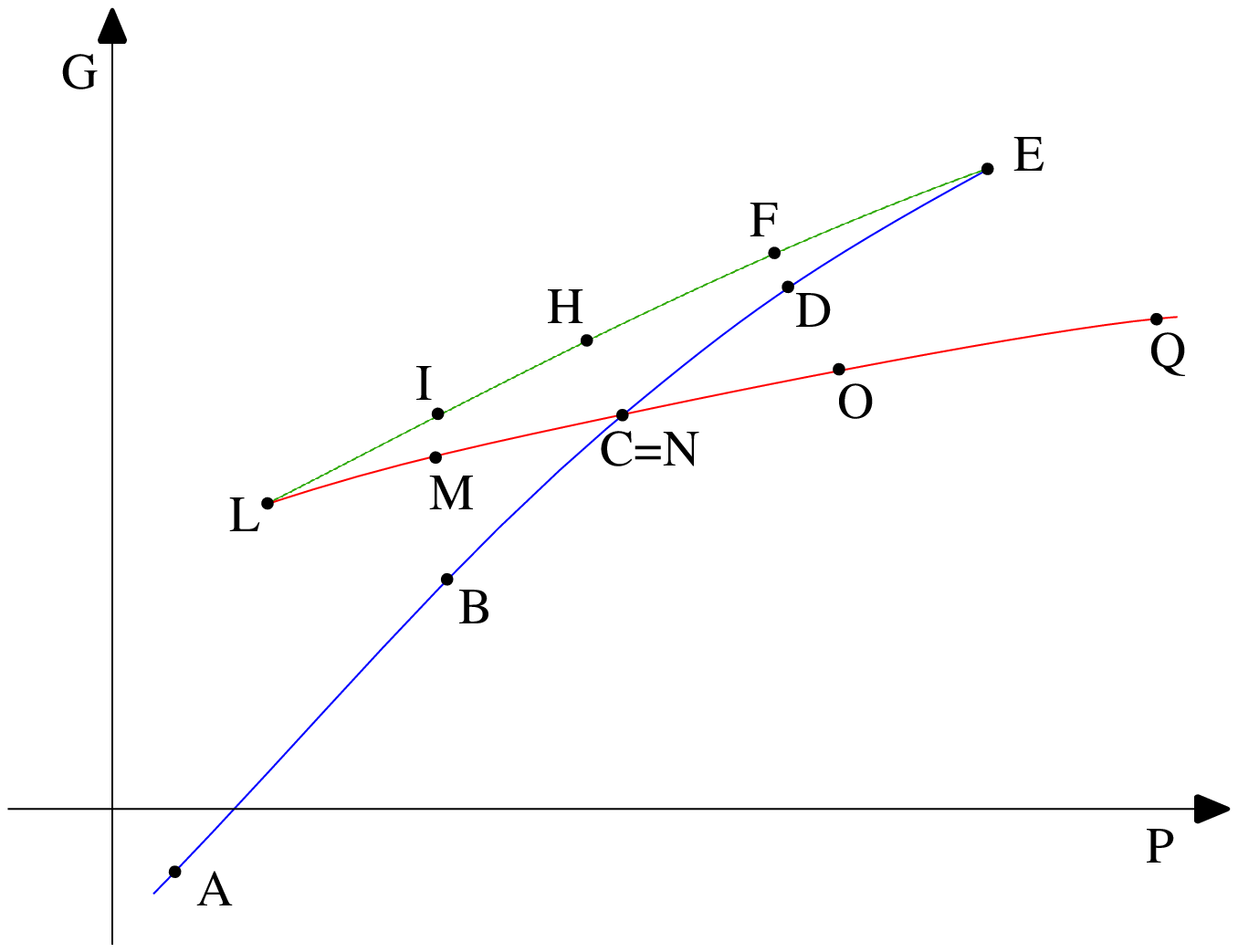}}
\caption{The pressure $P$ as a function of the volume $V$ (left panel)
and the Gibbs free energy $G$ as a function of the pressure $P$(right panel) for
the same isotherm curve.}
\end{figure}
Clearly this analogy works by linking $L(\r_0)$ to $P(V)$ as well as $E(L)$ to $G(P)$, such that $L\leftrightarrow P$, $\r_0 \leftrightarrow V$ and $E \leftrightarrow G$. 
On the VdW side, it is clear that a phase transition will be observed, if $P(V)$ has a local maximum. This can be achieved by tuning two scales, dictated by the constants $a$ and $b$, representing the interaction between the VdW gas particles and their non-zero size respectively. We have so far only explored a 1-dimensional parameter space. To increase our chances of success, we have to find another parameter to vary.
\subsection{Introducing $\r_*$} \label{flat}
One thing one can try is to let the source branes not quite reach 0 but only start at some point $\r_*$. This is easily achieved by multiplying the profile with a step function $\Theta(\rho-\rho_*)$ and performing the coordinate transformation $\r \rightarrow \r-\r_*$. So, for example, our bump profile becomes $S(\r)= \Theta(\rho-\rho_*)\tanh(2\r-2\r_*)^4 e^{-4(\r-\r_*)/3}$.

Numerically, one can simply solve the unflavored system between $0$ and $\r_*$ and then connect the system by using $P_{\textrm{unfl}}(\r_*)=P(\r_*)$ and $P'_{\textrm{unfl}}(\r_*)=P'(\r_*)$ as initial conditions for the flavored system. As shown in \cite{Conde:2011rg} this is justified if $S'(\r_*)=S"(\r_*)=0$ which is true in our cases. 

\begin{figure}[H] 
  \centering
    \includegraphics[width=0.5\textwidth]{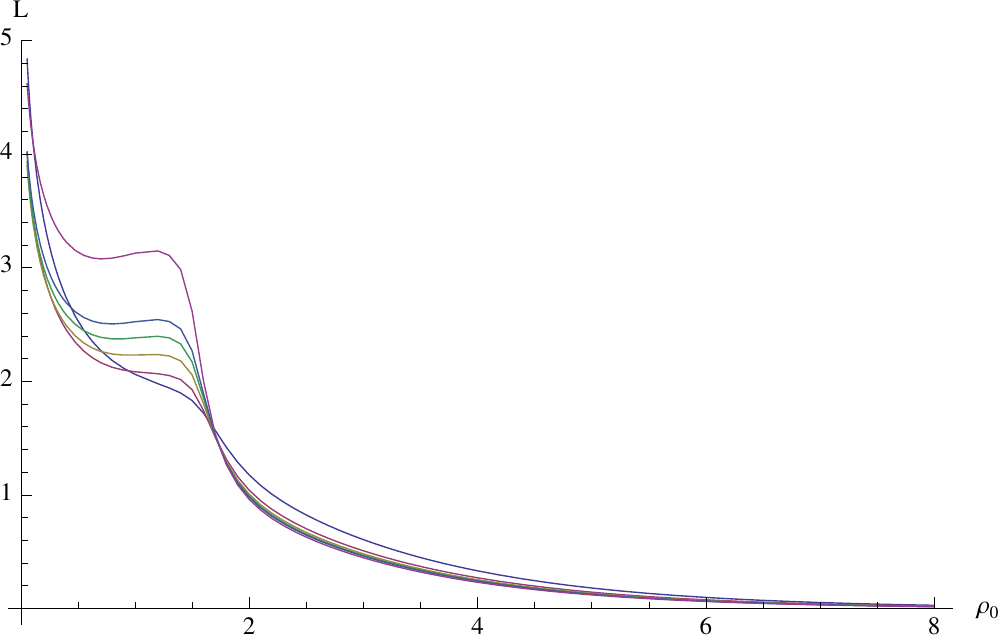}
    \caption{$L(\r_0)$ for the exact same system for $h_1=h_{1c}+5$ at the bottom, till $h_1=h_{1c}+45$ in increments of 10 and then at the top for $h_1=h_{1c}+100$.} \label{LCompare}
\end{figure}

Here, a phase transition can be found regardless of which profile has been used. When we studied the parameter space in detail, we found that as we increased $h_1$, $L(\r_0)$ began to flatten in the region just below $\r_*$ until a peak appeared that kept getting more pronounced. See Figure~\ref{LCompare}. This behaviour was so typical that we conjecture that given any acceptable profile and $\r_*>0$ a phase transition can be found for this system. As an example look at the following figures. 
\begin{figure}[H] 
  \centering
    \includegraphics[width=\textwidth]{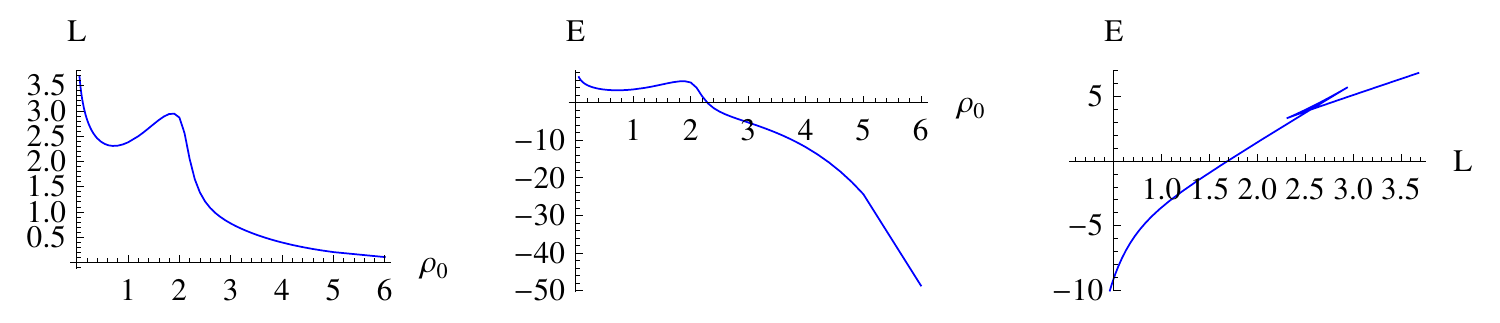}
    \caption{Here,$S(\r)=\frac{\tanh^4(2(\r-\r_*))}{(\sinh(4(\r-\r_*))-4(\r-\r_*))^\frac{1}{3}}$, $\Phi_0=0$, $N_c=1$, $N_f=1$, $h_1=27$ and $\r_*=2$}\label{r*}
\end{figure}
The position of the phase transition is not random and has a nice physical interpretation. Before the phase transition the string explores only the sourcefull region, while afterwards a majority of the string is located in the sourceless region $\r<\r_*$. Thus the phase transition appears as more and more of the string enters the sourceless part.

For a given $\r_*$, the value of $h_1$ that produces a phase transition is only bound below. This also has a nice physical picture associated to it. $h_1$ is directly related to $c+$, the expansion parameter of the UV asymptotes of $P$ \eqref{UV}. $c+$ in turn is related to the scale at which $P$ stops being linear and starts growing exponential. This in turn effects the gradient of $E(L)$ and thus the "speed" with which the string hits the sourceless region. For a more pictorial description, the reader may imagine a light ray getting refracted on a block of glass due to the fact that it is slowed down by the medium. The higher $h_1$, the sooner $P$ becomes exponential, and the "faster" the string hits the sourceless region. Above a certain threshold we observe phase transitions.
As the gradient of $L(\r_0)$ grows to $\infty$ as $\r_0 \rightarrow 0$, it is clear that the threshold for producing phase transitions, $h_{1t}$ also grows to $\infty$ as $\r_* \rightarrow 0$. This relationship between $\frac{dL(\r_0)}{d\r_0}$ and $h_{1t}$ exists as $h_1$ represents how much $L(\r_*)$ is lifted compared to close by points, and a steeper gradient requires more lift. See the figure above. For large $\r_0$, $L$ is already quite flat, so even small values of $h_1$ will cause a bump to appear. For small $\r_0$, $L$ is quite steep, so the initial increase in $h_1$ only flattens the profile. This is why we have not observed phase transitions just varying $h_1$ and effectively leaving $\r_* =0$.
Of course other parameters in the theory will also influence the development of phase transitions. Please refer to Appendix \ref{influence} for an exploration of them.

\section{Double phase transition}
In the last section we have described several ways to achieve phase transitions. This can be best understood through the variation of two scales, dictated by $h_1$, giving the "speed" of the increase of the potential with respect to separation length, and $\r_*$, the scale at which the $N_f$ sources kick in. In particular, a general characteristic of profiles has been identified that is key in producing phase transitions. We need a region where the source density rapidly changes to be able to observe them. This appears to be a universal requirement. Our interpretation of the reason for the occurrence of the phase transitions actually enables us to provide a testable prediction, that can be exploited to produce new phenomena.

Something that has been missing so far in the literature, is the possibility to find solutions with more than one phase transition. Below we list several types of profiles that allow us to generate these.

In this paper, we will restrict ourselves to profiles producing two phase transitions as proof of concept. In general though, these procedures seem to be able to produce an arbitrarily large amount of phase transitions if one so wishes.
\subsection{Double Bump and Tumbling profiles}
The first idea is to add several profiles of the form described in section~\ref{flavor} or \ref{bump}. Here, we used
\[
 S(\rho) =
  \begin{cases}
  \tanh^4(2(\r-\r_{*1}))+\tanh^4(2(\r-\r_{*2}))  & \text{if } \rho \geq \r_{*2} \\
 \tanh^4(2(\r-\r_{*1}))         & \text{otherwise}
  \end{cases}
\]
giving us a the following source profile.
\begin{figure}[H] 
  \centering
    \includegraphics[width=0.5\textwidth]{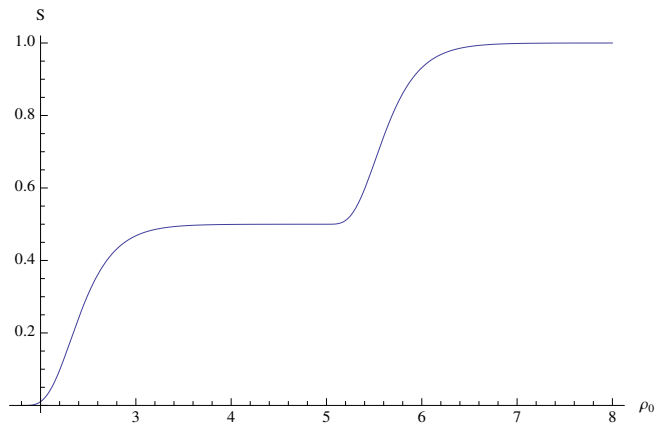}
    \caption{Here,  $\r_{*1}=1.8$ and $\r_{*2}=5$ }
\end{figure}
as well as
\[
 \hat{S}(\rho) =
  \begin{cases}
  \tanh^4(2(\r-\r_{*1})) e^\frac{-4(\r-\r_{*1})}{3}+\tanh^4(2(\r-\r_{*2})) e^\frac{-4(\r-\r_{*2})}{3} & \text{if } \rho \geq \r_{*2} \\
 \tanh^4(2(\r-\r_{*1})) e^\frac{-4(\r-\r_{*1})}{3}          & \text{otherwise}
  \end{cases}
\]
looking like
\begin{figure}[H] 
  \centering
    \includegraphics[width=0.5\textwidth]{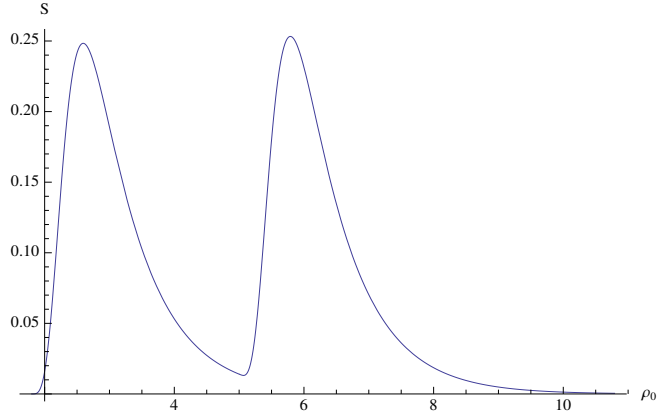}
    \caption{Here,  $\r_{*1}=1.8$ and $\r_{*2}=5$ }
\end{figure}
This leads to the following plots. For the Tumbling profile we have
\begin{figure}[H] 
  \centering
    \includegraphics[width=\textwidth]{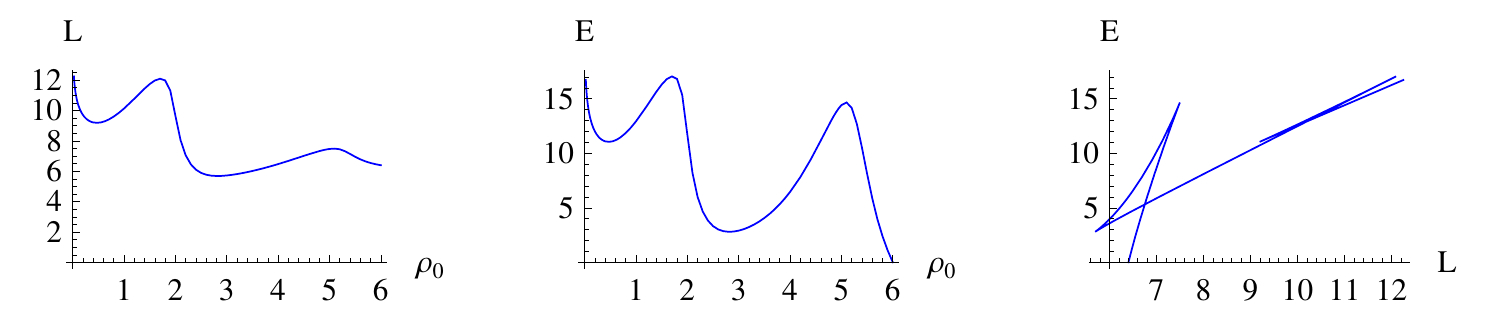}
    \caption{Here,  $h_{1}=47$ } \label{Tumbling}
\end{figure}
And for the Double Bump:
\begin{figure}[H] 
  \centering
    \includegraphics[width=\textwidth]{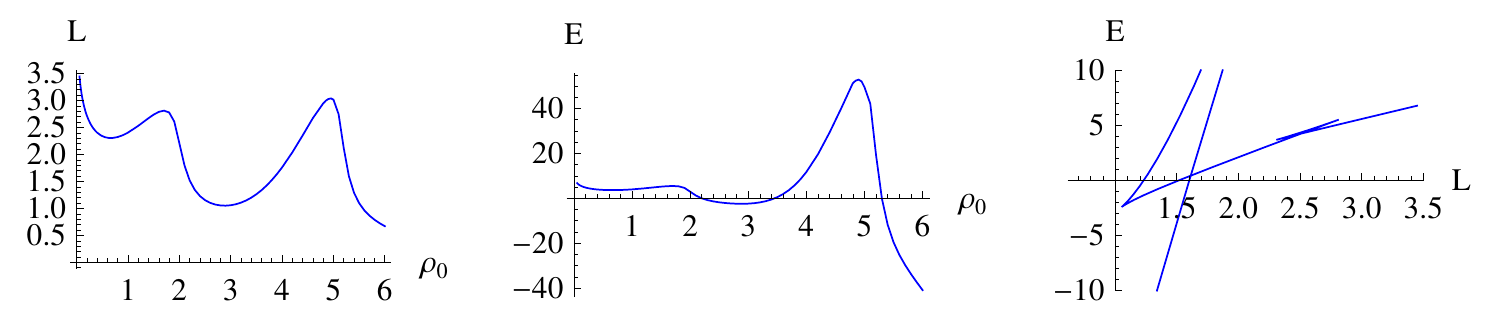}
    \caption{Here,  $h_{1}=47$ }
\end{figure}
Notice, that if the jump is from a region of high source density to a region with lower but non-zero source density, the smaller the difference, the harder it is to produce a phase transition, independent from the magnitude of the gradient of the jump. This can be seen in figure~\ref{Tumbling}, where in the left-hand diagram the second peak is a lot less pronounced than in the right one. 

Also the Tumbling profile in particular can be very interesting for Holographic Technicolor theories with Higgsing cascades. Here, each of the steps in the profile would correspond to a particular generation of quarks. This case is discussed in \cite{skeleton}.

Encouragingly, the central charge of the solutions of all profiles in this section, including plateau profiles are monotically increasing and thus show no sign of unphysical behaviour. To calculate the central charge, we used the same definition as in \cite{cascades}
\begin{equation}
c = \frac{\hat{h}^2e^{2\Phi+2h+2g+4k}}{8(\partial_\r \log[\sqrt{\hat{h}}e^{2\Phi+2h+2g+k}])^3}
\end{equation}
This leads to graphs of the following shape:
\begin{figure}[H] 
  \centering
    \includegraphics[width=0.5\textwidth]{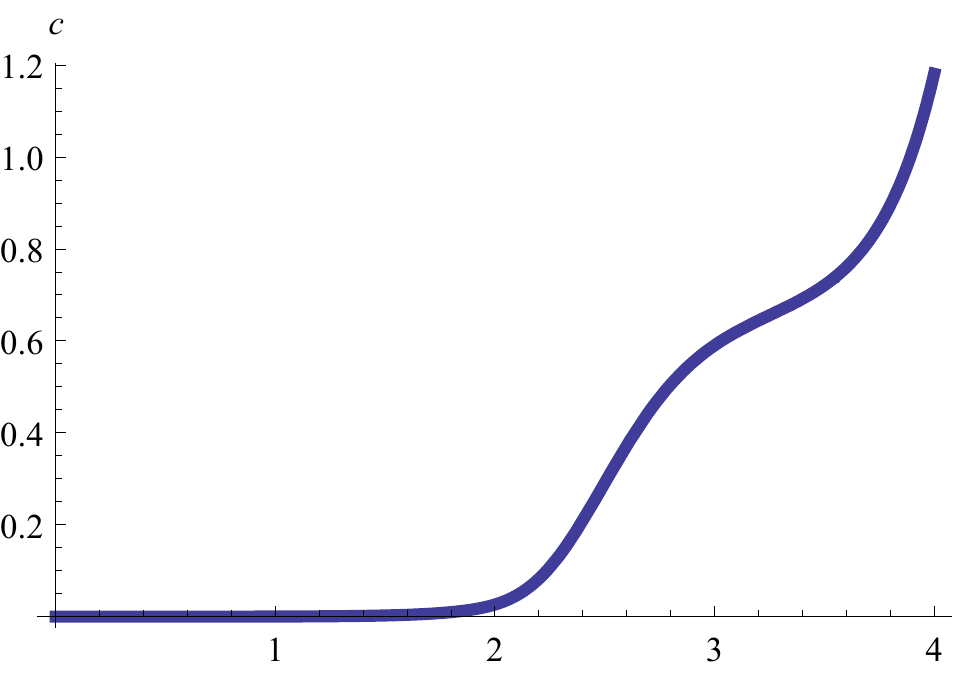}
    \caption{This is the central charge for the Tumbling case. For the other cases, the shape is similar, but c increases much faster}
\end{figure}
\subsection{Plateau profiles}
So far all the phase transitions observed come from the string moving from regions of high source density to regions of lower density. It would be interesting to see if this also works if we move from low to high density. Thus we would like to engineer profiles that quickly rise from $0$ to $1$, stay at $1$ for some time before quickly decreasing back to $0$. It is a major challenge to find such functions and so far none have been found that have a continuous first derivative. These can still be used, given that the discontinuity is well away from the regions of interest, such as the middle of the plateau, but will lead to an imperfection discussed below.
 
The easiest way to obtain profiles as described, is to use a profile that behaves like $S(\r)$ from section~\ref{flavor} and \ref{flat} and then mirror it around the axis $\r=c$, where $S(c)\approx 1$ and then forcing $S$ to be $0$ after it reached that value again. As an example we used
\[
 S(\rho) =
  \begin{cases}
	0  & \text{if } \rho \geq \r_{*}+4 \\
  \tanh^4(2(\r-\r_{*}-4))  & \text{if } \r_{*}+4 \geq \rho \geq \r_{*}+2 \\
 \tanh^4(2(\r-\r_{*}))         & \text{otherwise}
  \end{cases}
\]
resulting in
\begin{figure}[H] 
  \centering
    \includegraphics[width=0.5\textwidth]{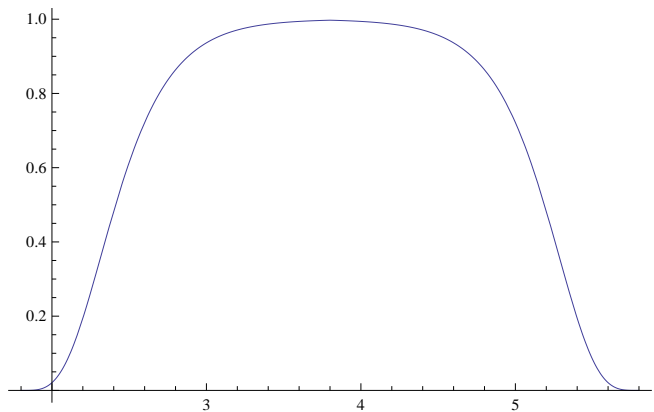}
    \caption{Here, $N_f=N_c=1$, $\r_*=1.8$ and $h_1=47$ }
\end{figure}
This leads to the following situation.
\begin{figure}[H] 
  \centering
    \includegraphics[width=\textwidth]{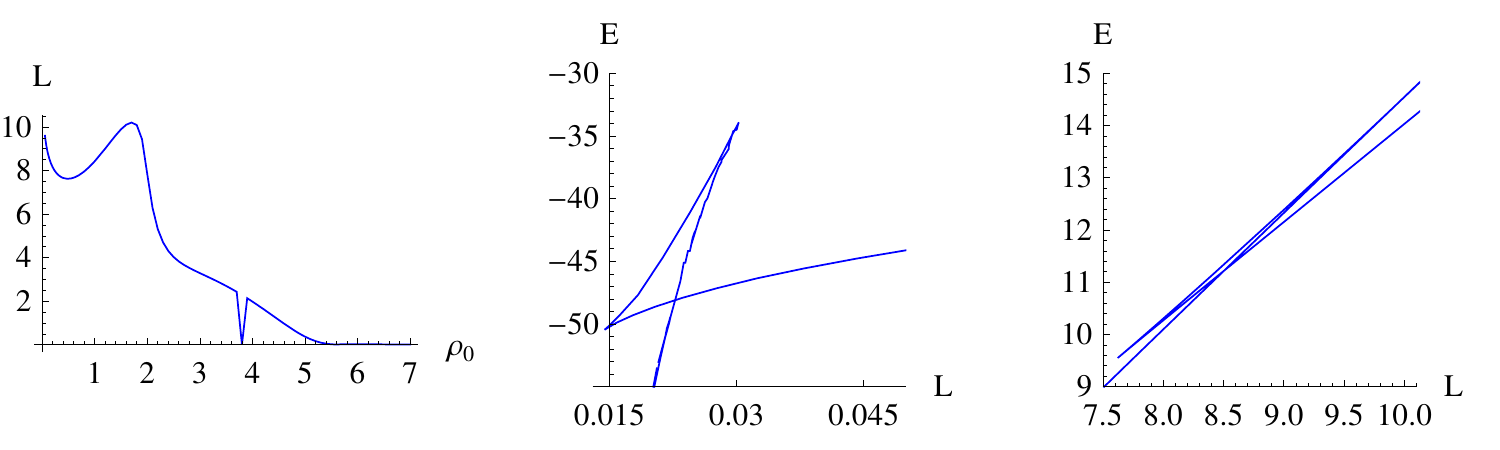}
    \caption{$L(\r_0)$ and both physical phase transitions }
\end{figure}
Please note that $E(L)$ has 3 phase transitions, the two physical ones shown above, and one unphysical that is due to the discontinuity of $S'$ at $\r=3.8$ at which the numerics produce the kink seen for $L(\r_0)$. Note that the background does not contain any singularities per se. It is just at the exact point of the discontinuity that the numerics break down. Thus this discontinuity is usually not noticeable and does not produce a third unphysical phase transition. In order to highlight this breakdown, $L(\r_0)$ has been evaluated only at discrete points and then interpolated inbetween. Furthermore $L(\r_0)$ was evaluated at exactly $3.8$ to produce an erronous data point. The purpose of this was to highlight the limitations of the numerical calculation, and illustrate the dangers of using profiles without a continuous derivative. Also due to the different scales and sizes at which the two phase transitions occur, we had to graph them separately to seem them clearly.

Thus we have strong evidence that phase transitions also occur when entering region of higher source density. This leads us to believe, that every bump profile actually produces two phase transitions per bump, that merge into another through an effect that is analogous to the effect that limits the angular resolution of lenses. Both bumps in $L(\r_0)$ or $E(\r_0)$ of, for example, figure~\ref{r*} are so close together that they merge into one.

\section{Conclusions}
The basic idea of this paper was to study the details of Wilson loops of a novel class of solutions \cite{cascades}, that generalise the KS \cite{KS} and baryonic branch \cite{BGMPZ} solutions, that, by the addition of sources, move the QFT to the mesonic branch..

This was done constructively, by starting from a relatively plain wrapped branes model and slowly adding all the necessary features, keeping track of the properties of the Wilson loops of the theory at each stage.

Afterwards we discussed the possibility of creating a phase transition by tuning two scales dictated by $h_1$ and $\r_*$.  This construction of phase transitions has been very successful and yielded a positive outcome in every case. We also managed to find a reason why these transitions occur.

Lastly, the possibility to create several phase transitions has been discussed, and several features a profile needs to produce these have been examined in detail, yielding a double phase transition in each case. In particular, a profile that might become useful for some extended holographic technicolor theories has been described. 

My procedures should also be applicable to the 2+1 d equivalent that has been discussed in \cite{Niall}.

Also please note that while confinement has not been directly mentioned, it is generally understood that the theories discussed are confining. \cite{Sonnenschein:1999if} gives several conditions for $f$ and $g$ that each guarantee confinement and it is straightforward to check that in all cases at least one of them is fulfilled. 

An interesting open question would be a detailed mathematical derivation of the exact requirements needed to obtain a phase transition. We may have some results on this in the near future.

\section*{Acknowledgments}
I would like to thank several people for their support and guidance. Carlos N\'u\~nez, Daniel James Schofield, Niall Macpherson and Stephen Bennett all patiently helped me with all the questions that I could think of. Last but not least I thank Sebastian Halter for sharing some of his Mathematica code that has been very valuable.

My research is funded by the STFC.

\appendix
\section{Influencing the phase transition}\label{influence}
\subsection{Powers of $\tanh$ and $x$}
Of course other variables do also have an effect on the Wilson loop and the associated phase transition. 

For example, we can generalise the $\tanh$ component in the profiles discussed above from, $\tanh^4(2\r)$ to $\tanh^{2n}(2\r)$ where $n\geq 2$ is an integer.  This has the conceptual imperfection though, that it pushes the flavour correction in the IR expansion of $P$ to higher orders of $\r$ and thus effectively diminishes the effect of the flavour branes on the system more and more.

Also $x=\frac{N_f}{N_c}$ being a variable that affects many aspects of this system, also plays a role for how easy phase transitions can be found. As an example, here is $E(L)$ for systems equal to figure~\ref{r*}.
\begin{figure}[H] 
  \centering
    \includegraphics[width=\textwidth]{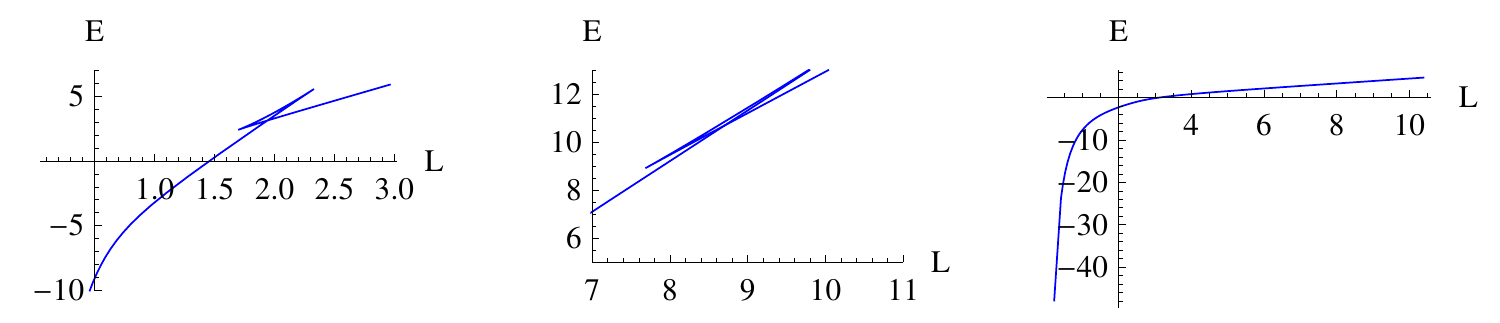}
    \caption{On the left $n=5$, in the middle $x=5$ and on the right $x=\frac{1}{5}$}
\end{figure}
Please note, that we can also find phase transitions for $x<1$ but higher values of $h_1$ have to be used.
\subsection{Different profiles}
Again, one could simply take a whole other class of profiles. Very helpful was
\[
 S(\rho) =
  \begin{cases}
   1-(\frac{\cosh{(4 \r q_1)}+\cosh{(4 \r q)}}{2}-1) e^{-4\rho} & \text{if } \rho \geq 4 \\
   \frac{2}{3}\frac{(\cosh{(4\rho)}-\cosh{(4 \r q)})^\frac{3}{2}}{(\cosh{(4 \r q_1)}-\cosh{(4 \r q)})\sqrt{\cosh{(4\rho)}-1}}            & \text{if } \rho \leq \r q_1 \\
     \frac{2}{3}\frac{(\cosh{(4\rho)}-\cosh{(4 \r q)})^\frac{3}{2}-(\cosh{(4\rho)}-\cosh{(4 \r q_1)})^\frac{3}{2}}{(\cosh{(4 \r q_1)}-\cosh{(4 \r q)})\sqrt{\cosh{(4\rho)}-1}}           & \text{otherwise}
  \end{cases}
\]
based on the Flat Measure described in \cite{Conde:2011rg}. Here, analytical conditions are found, that the profile must fulfil, to produce desirable back reactions. These are then solved in a lengthy but logical procedure whose details are mentioned in the paper cited. If one wishes to have a decaying profile one could use $\hat{S}(\r)=S(\r)e^{-\frac{4}{3}\r}$. These profiles have the following form
\begin{figure}[H] 
  \centering
    \includegraphics[width=0.5\textwidth]{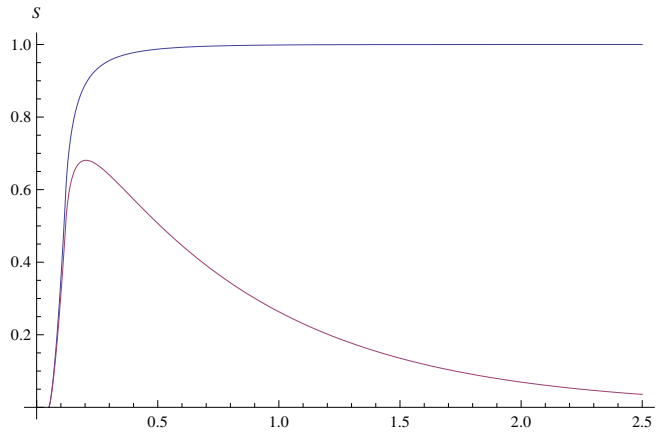}
    \caption{$S(\r)$ is blue while $\hat{S}(\r)$ is red}
\end{figure}
Both profiles lead to a phase transition. For $S(\r)$ we have
\begin{figure}[H] 
  \centering
    \includegraphics[width=\textwidth]{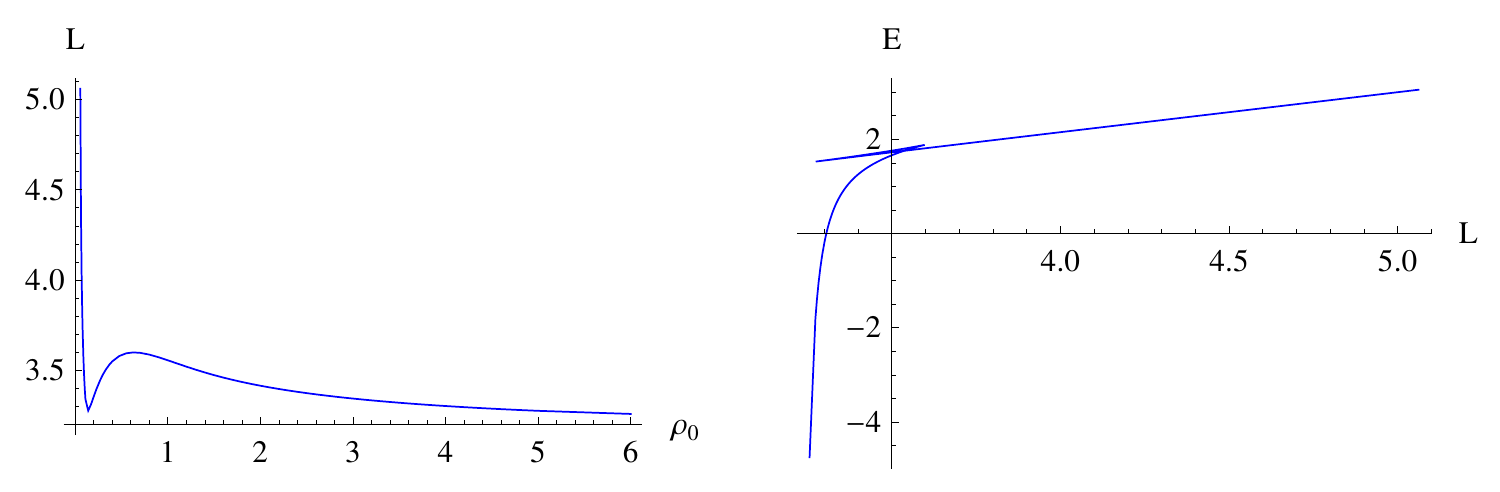}
    \caption{$N_c=N_f=1$ and $h_1=3.9581$}
\end{figure}
And for $\hat{S}(\r)$ we get:
\begin{figure}[H] 
  \centering
    \includegraphics[width=\textwidth]{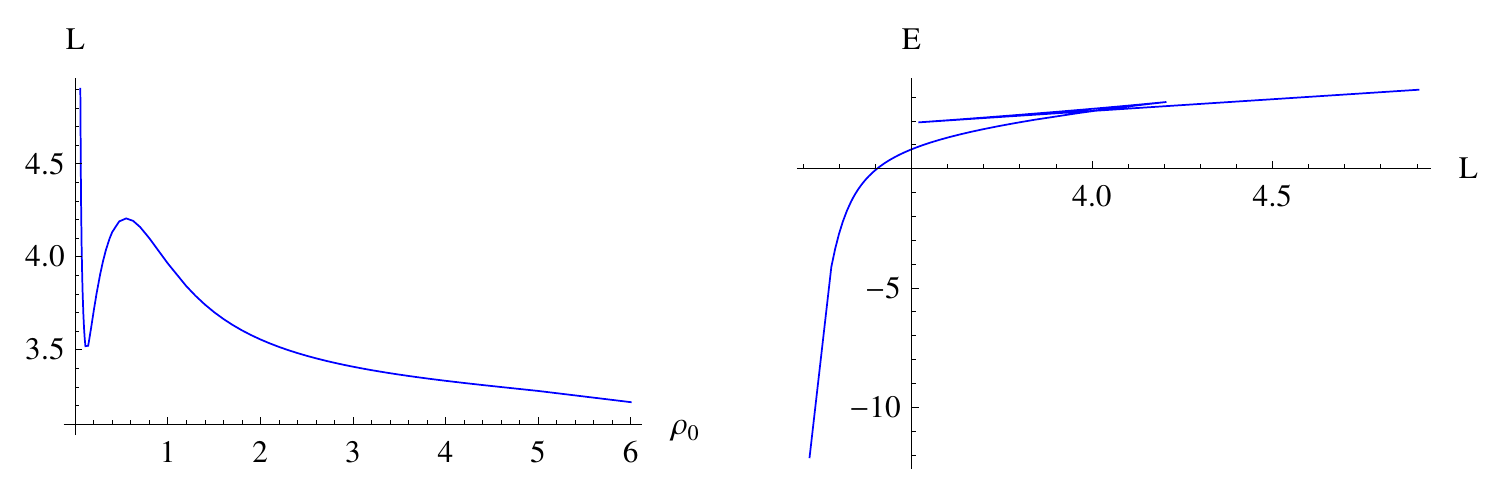}
    \caption{$N_c=N_f=1$ and $h_1=3.25291$}
\end{figure}
$\r q_1$ is related to $\r q$ via
\begin{equation}
\r q_1 = \frac{1}{4} \textrm{arcosh}(\d+\cosh(4\r q))
\end{equation}
As $\d$ decreases, the width of the brane distribution becomes smaller, The mass of the heaviest quark decreases and phase transitions are produced. Despite the large amount of control such types of profiles offer, one should be aware, that their realization is not completely clean, in the sense that $S'$ is not continuous everywhere, which can lead to problems as we will see in the following sections.
\section{Curiosities}
Here, some odd findings will be collected, that have been discovered during the analysis of the Wilson loops of the various Maldacena-Nunez solutions.
 
Firstly, it might be worth mentioning that, after the addition of flavor sources without as done in the beginning of section \ref{flavor}, we can produce a phase transition like behaviour in the lower, physical branch of $E(L)$ plot:
\begin{figure}[H] 
  \centering
    \includegraphics[width=\textwidth]{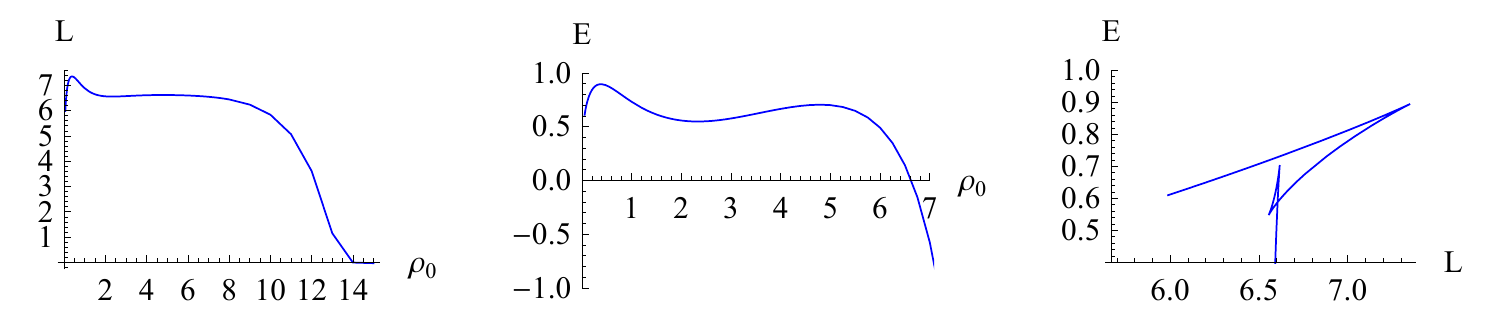}
    \caption{$\Phi_0=0$, $N_c=5$, $N_f=2$ and $h_1=29$}
\end{figure}
.

Another quite surprising fact that was discovered during the search for phase transitions is that while most profiles require $h_1 \gg h_{1c}$ to show transition, profiles of the type  $S(\r)=\tanh^4(2(\r-\r_*))$ do not. For $h_1 \approx h_{1c}$, $P(\r)$ will stay linear for quite some time before exhibiting exponential behaviour. This can be seen in Figure~\ref{P}. In such cases, $L(\r_0)$ will not immediately approach $0$, but first approach a non-zero value $\a$ before, and $\b$ after the transition of $P$'s behaviour. Depending on several factors, such as the power of $\tanh$, we can force $\b>\a$, leading to a very curious phase transition as shown in the following diagram.
\begin{figure}[H] 
  \centering
    \includegraphics[width=\textwidth]{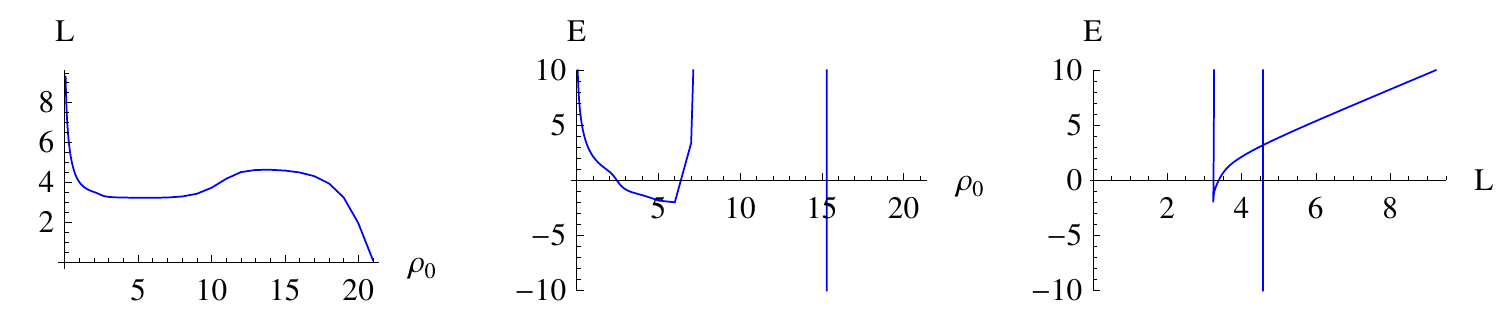}
    \caption{Here, $S(\r)=\tanh^4(2(\r-\r*))$,$N_c=N_f=1$, $h_1=\frac{1000519}{500000}$, $\r_*=2$ }
\end{figure}
Please note that beside the phase transition the overall shape of $E(L)$ is equivalent to the expected, but one would have to plot it approximately between $-10^7$ and $10$ to see it. Similarly $E(\r_*)$ peaks at about $10^5$ before descending rapidly.

\end{document}